\documentclass[twocolumn,footinbib,pre]{revtex4}

\usepackage[final]{graphics}
\usepackage{amssymb}
\usepackage{amsmath}
\usepackage{latexsym}

\begin{document}


\title{Competition between condensation of monovalent and multivalent
ions in DNA aggregation}
\author{Yoram Burak}
\email{yorambu@post.tau.ac.il}
\author{Gil Ariel}
\author{David Andelman}
\affiliation{School of Physics and Astronomy \\
Raymond and Beverly Sackler Faculty of Exact Sciences \\
Tel Aviv University, Tel Aviv 69978, Israel}
\date{March 21, 2004}

\begin{abstract}
We discuss the distribution of ions around highly charged PEs
when there is competition between monovalent and
multivalent ions,
pointing out that in this case the number of condensed ions is
sensitive to short-range interactions, salt, and
model-dependent approximations. This sensitivity is discussed in
the context of recent experiments on DNA aggregation,
induced by multivalent counterions such as spermine and spermidine.
\end{abstract}

\maketitle

\section*{\hspace{-0.71cm} 1. \ \ INTRODUCTION}

Despite extensive theoretical and experimental research,
polyelectrolyte (PE) solutions are
relatively poorly understood compared to their neutral
counterparts \cite{BarratJoanny}. The main difficulty in their
theoretical treatment arises from
the long-range nature of electrostatic interactions
between the charged groups along the PEs.
Another major difficulty arises in highly charged PEs due
to their coupling with the surrounding ionic solution, which
is difficult to treat theoretically,
since one cannot simply trace over the ionic degrees
of freedom via the linearized Debye-H\"{u}ckel theory.

In this paper we address
the distribution of ions near
highly charged PEs,
concentrating on the case where more
than one counterion species is present
in the solution.
We point out that
the number of condensed ions is then highly sensitive to
various parameters such as
short range interactions, salt concentration,
and model-dependent approximations - even at low
concentrations of salt.
In contrast, in solutions with only one type of counterion
these parameters are important at high
salt concentrations, \textit{e.g.}, in the ion-dependent solubility
of proteins; at lower salt concentrations,
typically up to 100 mM, their influence on ion
condensation is weak.

After illustrating the above points using a simple example
(Sec. 2), we discuss the competition between monovalent and
multivalent ions in the context of DNA aggregation (Sec. 3),
concentrating
in particular on the role played by short-range interactions
in the dilute (non-aggregated) phase. In Sec. 4 we
discuss qualitatively the dependence on salt concentration
of the number of condensed ions. For this purpose we
use a simplified two-phase model similar to
Manning's model.

\section*{\hspace{-0.71cm} 2. \ \ CONDENSATION ON A SINGLE POLYELECTROLYTE CHAIN}

Let us consider first the distribution of ions near
a single PE, \textit{i.e.}, taking the limit of
infinite dilution for the PE solution. Let us assume
also that the PE is uniformly charged with a charge
per unit length equal to $\rho$.
Suppose first that
there is only one type of counterions in the system.

When there is no salt in the solution only some of
the ions remain bound to the PE, while the others
escape to infinity. The number of bound ions, per unit
length, is given by the well-known formula obtained
from Manning condensation theory \cite{Manning1}:
\begin{equation}
\rho_b = \left\{
\begin{array}{ll}
\frac{1}{z}\left(\rho-\rho^* \right) & , \ \ \ \rho>\rho^*\\
0 & , \ \ \ \rho<\rho^* \\
\end{array}
\right.
\label{manning}
\end{equation}
where $\rho^*=1/(zl_B)$ and $l_B=e^2/(\varepsilon k_B T)$ is
the Bjerrum length, $z$ is the counterion valency,
$e$ is the unit charge, $\varepsilon$ is
the dielectric constant of the solvent and $k_B T$ is the
thermal energy. When $\rho > \rho^*$ the
condensed ions partially neutralize the PE such that
its effective charge per unit length is equal to
$\rho^*$. This result is a consequence of the interplay
between entropy and electrostatic energy at large ion-PE
distances. Hence short range interactions (ion-ion or ion-PE)
are not expected to modify the number of condensed ions,
although they may influence the distribution of the ions
within the condensed layer.

\begin{figure*}
\scalebox{0.40}{\includegraphics{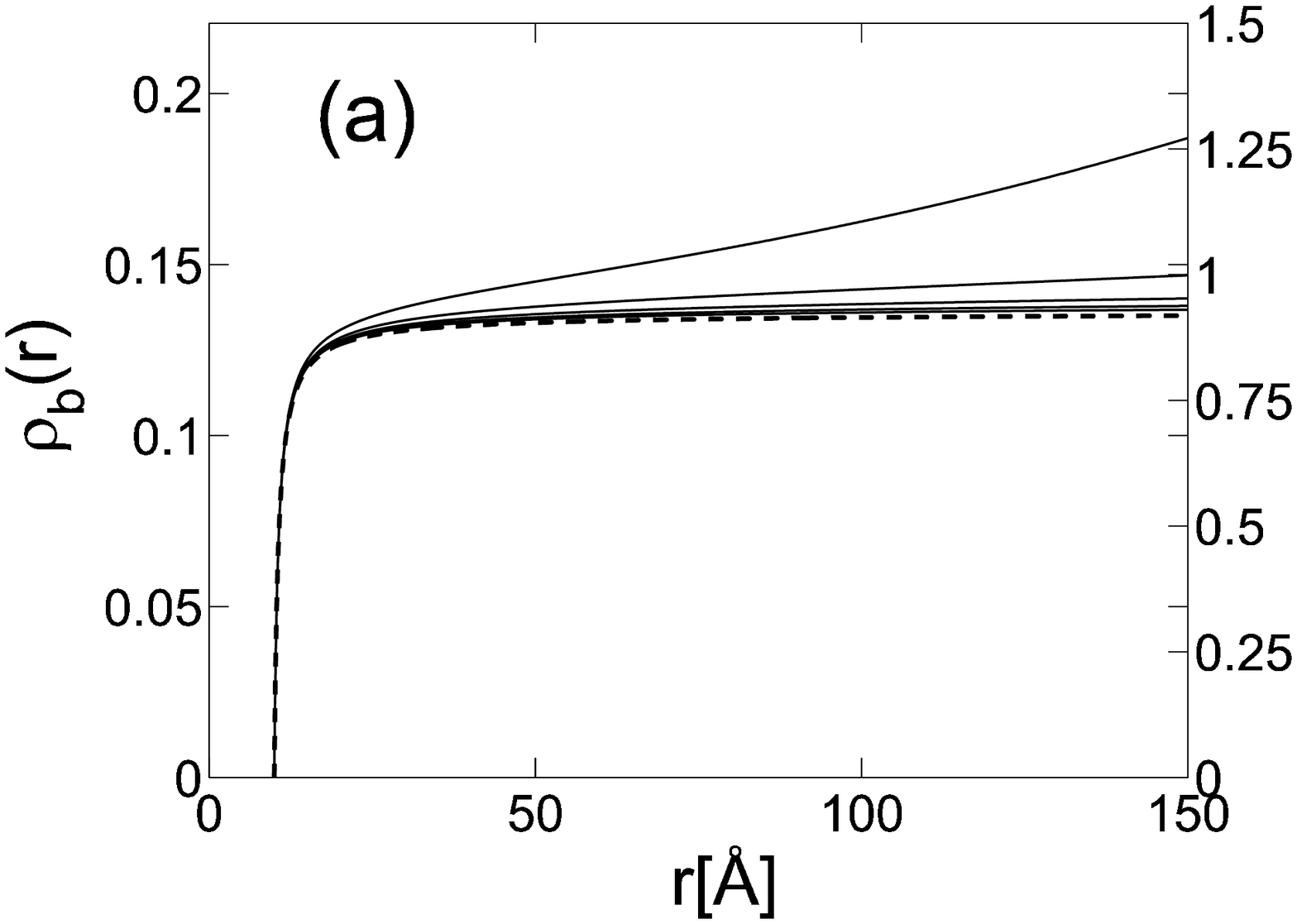}}
\hspace{0.3cm}
\scalebox{0.40}{\includegraphics{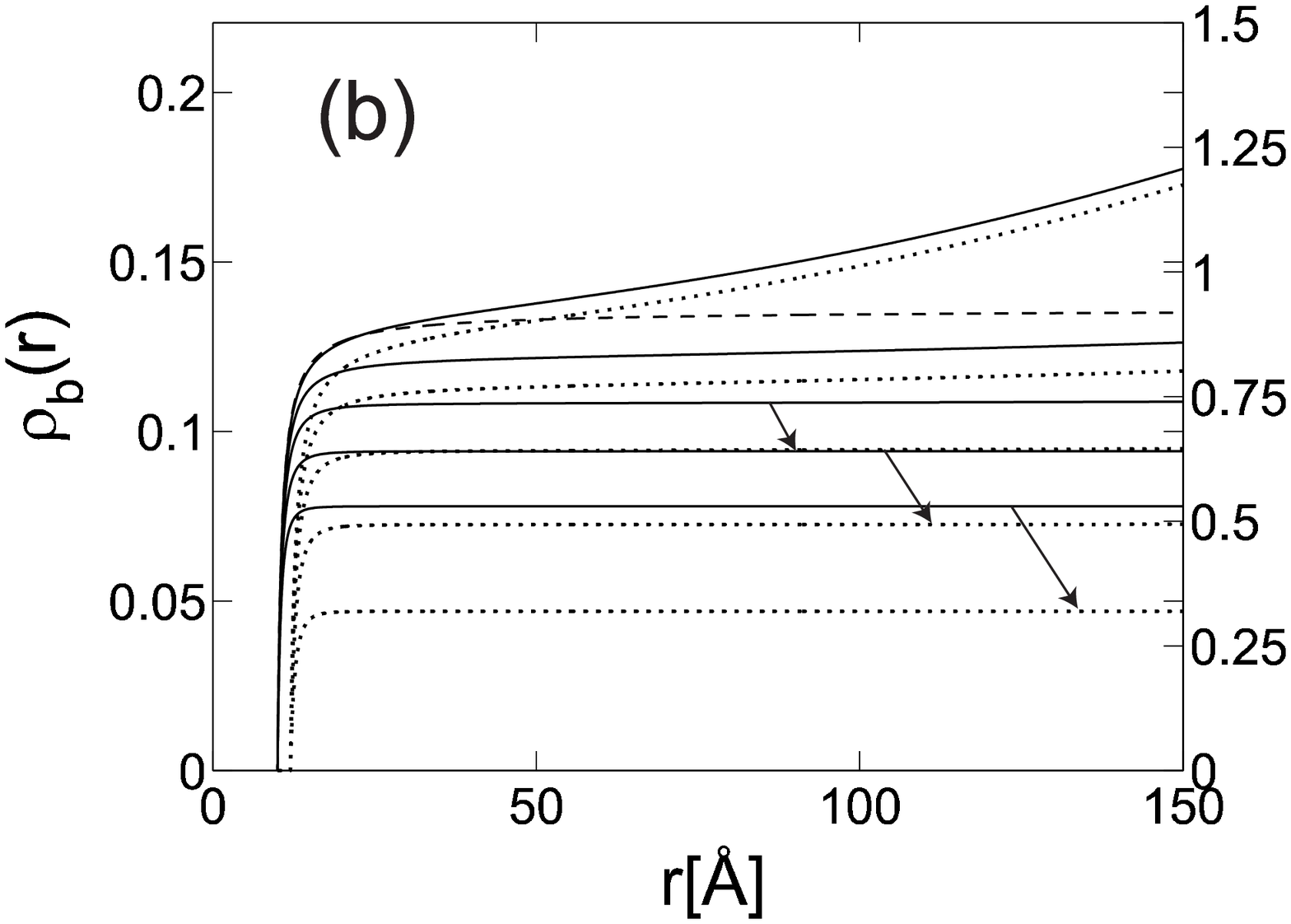}}
\caption{Accumulated number of 4-valent counterions, $\rho_b(r)$, up to
a distance $r$ from DNA,
modeled as a uniformly charged cylinder of
radius 10\,\AA: In (a), with only 4:1 salt of concentrations
0.01, 0.1, 1, 10, and 100\,mM (solid lines). The limiting case
of no salt (only 4-valent counterions) is also shown (dashed line).
In (b) 10\,mM of 1:1 salt is also present in the solution
(solid lines). Dotted lines
correspond to larger 4-valent counterions, having a distance of
closest approach of 12\,\AA \ to the DNA (the arrows connect
solid and dotted lines corresponding to the same 4:1 salt
concentrations of 0.01, 0.1, and 1\,mM).
The dashed line is the
same as in part (a). The right axis in
both figures shows
$a z \rho_b(r)$, the part of DNA charge compensated by the multivalent
ions. In both parts the ion density profiles are calculated
numerically using the Poisson-Boltzmann (PB) equation.
}
\end{figure*}

The number of ions in the vicinity of the polymer
is also insensitive to salt at low and moderate
concentrations \cite{Manning2},
as illustrated in Fig.~1\,a.
The figure shows the accumulated
number of multivalent ions ($z$=4)
per unit length up to a distance $r$ from
a PE having roughly the parameters of DNA
(charge per unit length equal to 1/(1.7 \AA)
and a radius of 10 \AA),
as a function of $r$. The ion distribution is modeled
using mean field theory and calculated using
the Poisson-Boltzmann (PB) equation. This is done for simplicity,
while in fact correlation effects beyond mean field are important
in this case since the ions are multivalent.
The dashed line shows the distribution when
there is no salt in the solution; as $r$ increases
the accumulated number of ions $\rho_b(r)$
approaches a constant, which is the number of condensed ions
per unit length predicted in Eq.~(\ref{manning}). When 4:1 salt is
added to the solution at increasing concentrations of
$c_z$ = 0.01, 0.1 1 and 10\,mM
(solid lines) the number of counterions
close to the polymer is almost unaffected. A significant
effect is seen only with $c_z = $ 100\,mM.

Short-range interactions have no effect on the number of condensed
ions in the limit of zero salt concentrations, and their
effect remains small at low salt concentrations. For example,
the effect of ion-PE dispersion forces
was recently estimated for condensation of monovalent ions
on DNA,
using the Poisson-Boltzmann equation and the Hamaker
approximation for the ion-PE dispersion
interaction \cite{Ninham02}. The effect was considerable at 1\,M
salt concentration, where electrostatic interactions are highly
screened. However below 100\,mM, even with
rather strong dispersion interactions, in the order of
four times the thermal energy close to
contact, the effect on ion condensation was small.

Returning to our numerical example, let us consider
the situation when there is more than
one type of counterion in the solution.
Suppose that the
solution contains monovalent (1:1) salt of concentration
$c_1$, multivalent (z:1) salt of concentration $c_z$, and
assume for simplicity that there is only one species of
monovalent co-ions. The salt concentrations $c_1$ and
$c_z$ then determine the distribution of ions
around the PE -- such that far away from the PE
concentrations of monovalent counterions,
multivalent counterions and co-ions decay
to $c_1$, $c_z$ and $c_1+z c_z$, respectively.

Figure 1\,b shows results for monovalent salt of
concentration $c_1$ = 10 mM and the same
4:1 salt concentrations as in part a (solid lines).
For each value of $c_z$ there is a distinct number of condensed
multivalent ions, which depends on
the multivalent salt concentration. This number
is very different from the
Manning prediction, Eq.~(\ref{manning}), for a single counterion
species (dashed line). The number of condensed multivalent
ions is now determined not only by a balance of entropy
and electrostatics, but also from
the competition with the monovalent ions. Furthermore,
this competition can be influenced by
ion-specific short range interactions
\cite{RouzinaBloomfield2}
leading to a strong
influence on the number of condensed ions.
As a simple example, the dotted lines show results for
multivalent ions that are slightly larger than the monovalent ions,
having a radius of closest approach to the PE that
is larger by 2\,\AA \ from that of the monovalent ones, leading
to a considerable decrease in the condensation of multivalent ions.
Compared to the case of identical ion sizes,
the 4:1 salt concentration has to be increased by roughly
an order of magnitude in order to have the same number of
condensed multivalent ions.

\section*{\hspace{-0.71cm} 3. \ \ COUNTERION COMPETITION IN DNA AGGREGATION}

\begin{table*}
\label{table5}
\centerline{
\begin{tabular*}{\linewidth}{l@{\extracolsep{\fill}}llll}
\hline
$c_1$[mM] & $c_z^*$[mM] & $a \rho_z^*$ & $a\rho_z$ (PB)
& $a\rho_z$ (SR) \\
\hline
$2$  & $\ \ \ \ \ 0 \pm 0.0003$    & $0.194 \pm 0.020$ & $0.186 \pm 0.005$
& $0.191 \pm 0.006$\\
$13$ & $0.011 \pm 0.002$ & $0.191 \pm 0.013$ & $0.178 \pm 0.002$
& $0.172 \pm 0.003$ \\
$23$ & $0.031 \pm 0.005$ & $0.173 \pm 0.025$ & $0.172 \pm 0.002$
& $0.163 \pm 0.003$\\
$88$ & $\ \,0.52 \pm 0.05$   & $0.135 \pm 0.026$ & $0.164 \pm 0.002$
& $0.149 \pm 0.003$\\
\hline
\end{tabular*}
}
\caption{Excess of 4-valent ions near DNA extracted from DNA
aggregation experiments,
compared with calculated values using Poisson-Boltzmann theory (PB)
(treating the DNA as a cylinder of radius 10\,\AA)
and Poisson-Boltzmann theory with short range interactions (SR).}
\end{table*}

In Ref.~\cite{Raspaud} DNA aggregation, induced by  multivalent ions,
was studied experimentally. The conditions for aggregation were
mapped with varying concentrations of monovalent salt,
multivalent salt, and DNA.
These experiments provide interesting evidence for the role
played by competition between different ion species.
We will concentrate on experiments
that were done in solutions containing the following ingredients:
short (150 base pair) DNA chains of concentration
$c_{\rm DNA}$ (base pairs per unit volume), spermine (4+)
of concentration $c_z$ and monovalent salt of concentration
$c_1$. At sufficiently high concentrations the multivalent
salt mediates an attractive
interaction between DNA chains, leading to aggregation
of the DNA and its precipitation from the solution.
For constant $c_1$ and with increase of $c_z$ DNA
starts to precipitate
when $c_z$ crosses a threshold value which we denote
as $c_{z,{\rm aggr}}$. The dependence of this threshold
on $c_1$ and $c_{\rm DNA}$ was measured in \cite{Raspaud}
in great detail.

At the onset of aggregation all the DNA is still
solubilized. The partitioning of multivalent ions
into free and condensed ions is thus
governed by their distribution around isolated DNA chains.
Within the relevant experimental parameters, ion
density profiles associated with different chains are decoupled
from each other due to screening by salt. As a result,
a linear dependence
of  $c_{z,{\rm aggr}}$ on $c_{\rm DNA}$
is expected theoretically \cite{BAA03}:
\begin{equation}
c_{z,{\rm aggr}} = c_z^* + a \rho_z^* c_{\rm DNA}.
\label{onset}
\end{equation}

The first term on the right hand side, $c_z^*$,
is the concentration of multivalent ions
far away from the DNA chains.
This quantity plays the role of the salt concentration,
which determines the distribution of ions near the DNA chains. The second
term is the contribution of condensed ions to the volume-averaged
concentration of multivalent ions. The coefficient
multiplying $c_{\rm DNA}$, $a\rho_z^*$, is
the excess of multivalent ions per DNA base, where $a$ = 1.7\,\AA \
is the monomer length (1$e$ per $a$ on the chain)
and $\rho_z^*$ is a spacial integral
of the local excess of multivalent ions:
\begin{equation}
\rho_z^* = 2\pi \int {\rm d}r\, r\left[c_z(r)-c_z^*\right]
\end{equation}
where $c_z(r)$ is the local concentration. We will assume that
$c_z^*$ and $\rho_z^*$ do not depend on
$c_{\rm DNA}$, which is justified for
sufficiently long chains for which translational entropy
can be neglected \cite{BAA03}. Indeed,
the experimental measurements of $c_{z,{\rm aggr}}$, as function
of $c_{\rm DNA}$, fall on straight lines (within experimental
error bars) across
several orders of magnitude of DNA and spermine concentrations. The
coefficients $c_z^*$ and $a \rho_z^*$ can be extracted from
the experimental data \cite{BAA03}
and are reproduced in Table I.

\subsection*{\hspace{-0.67cm} 3.1. Comparison with PB theory}

Table I lists the concentrations of monovalent salt, $c_1$,
multivalent salt $c_z^*$ (extracted from the linear fit to
Eq.~(\ref{onset})),
and the excess of multivalent ions $a \rho_z^*$.
The last quantity is a measure for the number of
condensed multivalent ions at the onset of aggregation. Note that
it is of the same order of magnitude for all four monovalent
salt concentrations. However, $c_z^*$ varies by at least
four orders of magnitude. The large variation in $c_z^*$
is a consequence of competition between
monovalent and multivalent counterions: a relatively small increase
in monovalent salt concentration requires a large
addition of multivalent salt in order to keep
the number of condensed multivalent ions constant. This point
is further discussed in Sec. IV.

Table I provides simultaneous measurements of the salt
concentrations ($c_1$, $c_z^*$) and the excess of condensed multivalent
ions $a \rho_z^*$. Note that no particular model specifying the relation
between $c_1$, $c_z^*$ and $a \rho_z^*$ is assumed in Eq.~(\ref{onset})
and the latter two quantities are obtained independently from the linear fit.
This data
can be used to test any particular theory used to calculate
ion distributions around DNA. Such a comparison, with
Poisson-Boltzmann (PB) theory, is shown in the
fourth column of Table I:
for each pair of salt concentrations
($c_1$, $c_z^*$) the excess $a \rho_z$, as calculated from
PB theory, is compared with the experimental value of $a\rho_z^*$.

For the three smaller values of $c_1$ = 2, 13, and 23\,mM there is
a reasonable agreement with experiment (within the error bars).
However, for $c_1$ = 88\,mM there is a 30\% deviation. The overall
agreement with PB theory is surprisingly good, considering that
PB theory does not work so well for bulky multivalent ions.
Ion-ion correlations that are ignored in PB theory and are
important with multivalent ions, tend to increase the number of
bound multivalent counterions. Instead, for $c_1$ = 88\,mM, the
number of bound multivalent ions is decreased. We conclude that
ion correlations by themselves are not the main source for deviations
seen in Table~I, and short range interactions also play a prominent
role.

There are many types of short-range interactions that are not
taken into account in PB theory.
Spermine is a long, relatively narrow molecule, which
can approach DNA at close proximity, and even
penetrate the grooves at certain sites and orientations
\cite{Lyubartsev2}. On the other hand,
configurations that are close
enough to the DNA are accompanied by a loss of orientational
entropy. Other factors that modify the
interaction of spermine with DNA, compared to simplified
electrostatic models, include dispersion interactions,
specific ordering of charges on the spermine and DNA, and
arrangement of the surrounding water molecules.

Taking all the above parameters into account is beyond the scope of
this work. Instead we demonstrate, within the framework
of PB theory, that short-range interactions can
influence the competition between monovalent and multivalent ions,
and thereby affect the onset of aggregation in a similar way
to that seen it Table I.
As a simple example (with somewhat arbitrary parameters chosen
to demonstrate our point)
two short-range effects are added to the PB model.
We consider $4$-valent ions that are larger than the monovalent
ones. Hence the distance of closest approach to the DNA is
different for the two species. In this example these distances are
taken as $9\,\mbox{\AA}$ for the monovalent counterions and
$12\,\mbox{\AA}$ for the multivalent ones. In addition, we include
a short-range attraction between the multivalent ions and DNA:
multivalent ions gain $3\,k_B T$ if their distance from the DNA is
smaller than $15\,\mbox{\AA}$. Qualitatively these are two
competing effects. The first one (closer approach of monovalent
ions) slows down replacement of monovalent ions by multivalent
ions, while the second (short-range attraction) has the opposite
effect. The balance between the two effects is different for
different $c_1$ and $c_z^*$.

The last column of Table I
shows values of $a \rho_z$ calculated using the above
modified model. These values (SR) are shown next to the results of
the usual Poisson-Boltzmann theory (PB) and compared with the
experimental value of $a \rho_z^*$. For $c_1=2\,{\rm mM}$, $\rho_z$ is
almost the same in the two calculations. For $c_1=88\,{\rm mM}$
and $c_z^*=0.52\,{\rm mM}$, $\rho_z$ is considerably
decreased with the inclusion of short-range interactions, and is
closer to the experimental value. Any one of the two short-range
effects, by itself, results in a large discrepancy with experimental
data at low salt concentrations.

%
We believe that the importance of competing mechanisms for a long,
multivalent ion such as spermine go beyond the
simple modifications to PB described above. More refined
modifications include the loss of orientational entropy at close
proximity to the DNA. This effect creates a short-range repulsion,
whereas the correlation effect beyond mean-field is similar to a
short-range attraction.
Similar competing mechanisms were found in
simulation of spermidine ($3^+$) and NaCl in contact with DNA
\cite{Lyubartsev3}. In particular, for high salt concentrations
spermidine binding was considerably reduced compared to Poisson-Boltzmann
theory. In the computer simulation \cite{Lyubartsev3}
both molecular-specific
interactions, the geometrical shape of the constituents and ion-ion
correlations were taken into account. All these effects, and
especially the geometry of spermidine, which is similar to that
of spermine, were found to play an important role.

\section*{\hspace{-0.71cm} 4. \ \ TWO-PHASE MODEL FOR COMPETING SPECIES}

Two-phase models have been widely used to describe the
distribution of counterions around cylindrical macromolecules
\cite{Manning1,Oosawa}. In these models ions are considered as
either condensed or free. The condensed ions gain electrostatic
energy due to their proximity to the negatively charged chain but
lose entropy, since they are bound at a small cylindrical shell
around it. For systems with more than one type of counterion
Manning introduced the so-called two-variable theory
\cite{Manning3}, which is an extension of his previous model
\cite{Manning1,Manning2}. This model has been used to analyze
condensation (single molecule collapse) of DNA molecules induced
by spermine and spermidine
\cite{Wilson1,Wilson2}. In this section we present a similar
model, which differs from Manning's two-variable theory in
some details. Our main purpose is to explain the
large sensitivity of $c_z^*$ to changes in monovalent salt concentration.
As a by-product of our analysis we compare our two-phase model with
PB theory and Manning's two-variable theory.

\subsection*{\hspace{-0.67cm} 4.1. Model details and main equations}

Assume that the PE is confined within a finite cylindrical
cell of area A. The free energy is then written as follows:
\begin{eqnarray}
F & = & \rho_z\log\left(\lambda^3\frac{\rho_z}{A_c}\right)
      + \rho_1\log\left(\lambda^3\frac{\rho_1}{A_c}\right)
\nonumber \\
  & + & \rho_z^f\log\left(\lambda^3\frac{\rho_z^f}{A-A_c}\right)
      + \rho_1^f\log\left(\lambda^3\frac{\rho_1^f}{A-A_c}\right)
\nonumber \\
  & + & \frac{1}{2}(-\rho_{\rm DNA} +z\rho_z + \rho_1)\phi
\label{free_energy}
\end{eqnarray}
The first two terms are the entropy of bound multivalent
and monovalent counterions, where $\rho_z$ and  $\rho_1$ are
the number of condensed ions per unit length of the PE.
We assume that
condensed ions are bound on a cylindrical shell around the chain
and take its area, for simplicity, to be
\begin{equation}
A_c = \pi d^2
\end{equation}
where $d$ is the radius of the PE.
The length $\lambda$ is included in order to have a dimensionless
argument inside the logarithms, and can be chosen arbitrarily.

The next two terms are the entropy of free counterions.
The numbers per unit area of free multivalent ions, $\rho_z^f$,
and of free monovalent ions, $\rho_1^f$, are related
to the number of condensed ions since the total number of ions
within the cell is fixed:
\begin{eqnarray}
\rho_z^f & \equiv & (A-A_c) c_z^f  = A c_z  - \rho_z,
\nonumber \\
\rho_1^f & \equiv & (A-A_c) c_1^f  = A c_1 + \rho_{\rm DNA} - \rho_1.
\label{total_twophase}
\end{eqnarray}
where we introduced the concentrations of free counterions
$c_z^f$ and $c_1^f$.

Finally, the electrostatic energy is evaluated as if all
the bound ions are exactly at the cylinder rim, $r=d$,
and the linearized
Debye-H\"{u}ckel approximation is used
for the electrostatic potential at $r>d$. This leads to the
last term in Eq.~(\ref{free_energy}),
where $\phi$ is the electrostatic
potential at $r=d$, given by
\begin{equation}
\phi = -2 l_B (\rho_{\rm DNA}-z \rho_z - \rho_1)
       \frac{K_0(\kappa d)}{\kappa d K_1(\kappa d)}~,
\label{phi}
\end{equation}
where we assume that the outer cell radius is much larger than $d$
and $\kappa^{-1}$,
$\rho_{\rm DNA}$ is the number of unit charges per unit length of DNA,
$K_0$ and $K_1$ are zeroth and first order modified Bessel functions
of the first kind, respectively, and $\kappa^{-1}$ is the Debye length:
\begin{equation}
\kappa^2 = 4\pi l_B \left[2 c_1^f + z(z+1)c_z^f\right]
\label{kappa_twophase}
\end{equation}
%

The number of condensed monovalent and $z$-valent counterions
is found by minimizing the free energy with respect to
$\rho_1$ and $\rho_z$, yielding:
\begin{equation}
\log\left(\frac{\rho_z}{c_z^f A_c}\right)
 = -z \phi \ \ ; \ \
\log\left(\frac{\rho_1
}{c_1^f A_c}\right)
 = -\phi.
\label{ro1z}
\end{equation}
%
%

\subsection*{\hspace{-0.67cm} 4.2. Consequences for DNA aggregation}

We are interested in the qualitative dependence of
$c_z^*$ on $c_1$. Note that $c_z^*$ has the same role as
$c_z^f$ in the two-phase model. For the monovalent salt,
assuming that $c_1 > c_{\rm DNA}$,
$c_1^f$ can be replaced by $c_1$ \cite{BAA03}.
Equation (\ref{ro1z}) then
yields the following relation:
\begin{equation}
c_z^* = \frac{\rho_z^*}{A_c}\left(\frac{c_1 A_c}{\rho_1^*}\right)^z
\label{relation}
\end{equation}
where $\rho_1^*$ is the linear density of bound monovalent ions
at the onset of aggregation \cite{RouzinaBloomfield1}.
Qualitatively,
$\rho_1^*$ is the only ingredient that needs to be estimated in
this equation, since $c_1$ is controlled experimentally and $z \rho_z^*$
is of order one.

The main outcome of Eq.~(\ref{relation}) is that $c_z^*$
scales roughly as $(c_1)^z$. This explains the large variation of $c_z^*$
at different monovalent salt concentrations since $z=4$.
There are several sources for corrections to
this scaling. The first one is the dependence of $\rho_1^*$ on
$c_1$ and $\rho_z^*$.
A second source of corrections is the effect of
short-range interactions, which was discussed in the previous
section within PB theory. In addition,
Eq.~(\ref{relation}) involves all the approximations of the
two-phase model.

\subsection*{\hspace{-0.67cm} 4.3. Comparison with other models}

\begin{table}
\label{table4}
\centerline{
\begin{tabular*}{\linewidth}{l@{\extracolsep{\fill}}llll}
\hline
$c_1$ & $a\rho_z^*$ & $c_z^*$ (two-phase) & $c_z^*$ (PB) & $c_z^*$ (Manning) \\
\hline
$2$  & $0.194$ & $4.1 \times 10^{-4}$ & $4.3 \times 10^{-4}$ & $7.5 \times 10^{-7}$ \\
$13$ & $0.191$ & $1.0 \times 10^{-1}$ & $3.7 \times 10^{-2}$ & $3.4 \times 10^{-4}$\\
$23$ & $0.173$ & $7.4 \times 10^{-2}$ & $3.3 \times 10^{-2}$ & $4.1 \times 10^{-4}$\\
$88$ & $0.135$ & $3.9 \times 10^{-1}$ & $1.1 \times 10^{-1}$ & $4.8 \times 10^{-3}$\\
\hline
\end{tabular*}
}
\caption{Comparison of two-phase models (two-phase, Manning) with
PB theory. All the concentrations are in mM.}
\end{table}

We conclude this section by comparing the predictions of the
two-phase model with those of PB theory and Manning's two-variable
theory (see also \cite{BelloniDrifford,Wilson3}).
This is instructive due
to the wide use of two-phase models in the literature.
Table~II
lists the value of $c_z^*$ calculated with the two-phase model,
using the values of $c_1$
and $\rho_z^*$ of Table~I. The two-phase model can be seen to
agree qualitatively with PB theory. Quantitatively, their
predictions differ by a factor of up to four in the table.

Our two phase model differs from Manning's two variable theory
in some details. First, the area used in the expression for the
entropy of bound counterions is different. Second, the expression
for the electrostatic energy of bound ions is given in Manning's
theory by:
\begin{equation}
\phi = 2 l_B \left(\rho_{\rm DNA} - z\rho_z - \rho_1\right)
\log\left(1-{\rm e}^{-\kappa a}\right).
\label{phimann}
\end{equation}
Note that for small
$\kappa d$ the two forms in Eqs.~(\ref{phi}) and (\ref{phimann}) are
similar if $d$ is replaced by $a$, since:
\begin{equation}
\frac{K_0(\kappa d)}{\kappa d K_1(\kappa d)}
\simeq -\log\left(1-{\rm e}^{-\kappa d} \right).
\end{equation}

In the last column of Table~II we present the results
of Manning's two variable theory, in the version that was
used in Refs.~\cite{Wilson1,Wilson2,Wilson3}
(with different areas of condensation
for monovalent and multivalent counterions).
Compared to our two
phase model, deviations from PB theory are
larger, typically of approximately two orders of
magnitude.
Since both two-phase models
are quite similar to each other,
their different predictions demonstrate
the large sensitivity to model-dependent parameters.
In our opinion
such models are useful for obtaining
qualitative predictions, but should be used with great
care when quantitative predictions are required.

\section*{\hspace{-0.71cm} \ \ 5. SUMMARY}

In this paper we discussed competition between ions of different
valency in DNA aggregation, concentrating on DNA-counterion
complexes in the dilute (non-aggregate) phase.
Due to competition, the number of
condensed multivalent ions is highly sensitive to
salt concentration and to short-range,
ion-specific effects.
Simplified models that include only electrostatic
interactions are thus limited in their capability to
predict the conditions required for aggregation.
An important experimental evidence for
the importance of specific interactions is that
different multivalent ions
vary strongly in their ability to induce condensation or
aggregation of DNA, even if they have the same
valency \cite{DengBloomfield,Thomas}.
In addition to the role of specific interactions in the
dilute DNA phase, they also play a prominent role in the
aggregates \cite{Parsegian1,Parsegian3},
where the gap between neighboring
DNA chains is typically of order 10\,\AA \ \cite{Pelta}.

\begin{acknowledgments}
We are grateful to E. Raspaud and J.-L. Sikorav for numerous discussions.
Support from the U.S.-Israel
Binational Science Foundation (B.S.F.) under grant No. 98-00429
and the Israel Science Foundation under grant No.~210/02 is
gratefully acknowledged.
\end{acknowledgments}

\end{document}